\documentclass[9pt,twocolumn,twoside]{osajnl}
\journal{ao} % Choose journal (ao,jocn,josaa,josab,ol,optica,pr)

\setboolean{shortarticle}{false}
\usepackage{lineno}
\begin{document}

\title{Spectral shaping of an ultrafast modelocked Ytterbium fiber laser output through a passive intracavity optical filter; a simple and reliable route to sub-45 fs pulses}

\author[1]{Nicholas D. Cooper}
\author[1]{Uyen M. Ta}
\author[1,*]{Melanie A. R. Reber}

\affil[1]{Department of Chemistry, University of Georgia, 140 Cedar Street, Athens GA 30602, USA}

\affil[*]{Corresponding author: mreber@uga.edu}

\begin{abstract}

Here we investigate the use of passive intracavity optical filters for controlling the laser output spectrum of a polarization-mode-locked, ultrafast Ytterbium fiber laser. With strategic placement of the filter cutoff frequency, the overall lasing bandwidth can be increased or extended. Overall laser performance, including pulse compression and intensity noise, is investigated for both shortpass and longpass filters with a range of cutoff frequencies, tuned by rotating the filters. We demonstrate the use of an intracavity filter not only shapes the output spectra, it provides a route for overall broader bandwidths and shorter pulses in Yb:fiber lasers. These results demonstrate that spectral shaping is a useful tool to routinely achieve sub-45 fs pulse durations in Yb:fiber lasers.

\end{abstract}

\setboolean{displaycopyright}{false}

\maketitle

\section{Introduction}

Ultrashort pulses are desirable for a range of applications, including time-resolved spectroscopy, microscopy, and even optical machining. The design of lasers with increasingly shorter pulses is an active area of research even 40 years after the first ultrafast (sub-picosecond) laser was reported. Ti:Sapphire lasers, with a broad, flat gain bandwidth, are still the standard laser for generating sub 100-fs and even sub 10-fs ultrafast pulses; sub-10 fs lasers are now commercially available from multiple companies. However, Ti:sapphire lasers are energy intensive and environmentally sensitive, a result of the free-space layout and required high pump intensity. Rare earth-doped fiber lasers are an attractive alternative, since they are inherently more stable and require significantly less pump power. These lasers are flexible and versatile systems for the formation of ultrashort broadband pulses\cite{Fermann_JSTQE2009, Chong_RoP2015}, precision CW lasers, and even exotic structures like stabilized soliton pulses \cite{Song_ApplPhysRev2019}. The gain bandwidth of rare-earth doped fiber is not as large and flat as the Ti:Sapphire gain bandwidth, so it is generally challenging to obtain sub-100 fs pulses; they have not yet replaced Ti:Sapphire lasers in ultrafast laser labs. Ytterbium-doped fiber is still attractive as gain medium given the relatively broad (100-150 nm) gain bandwidth, and small quantum defect for efficient pumping\cite{Pashotta_JQE1997}. 

The shortest Ytterbium:fiber laser pulses have been achieved by managing the intracavity dispersion. Ilday et al achieved 36 fs pulses by optimizing the dispersion in the laser, which was about $30\%$ higher than the transform limit\cite{ilday_OptExpress2003}. They used a grating compressor inside the laser cavity to compensate for the intracavity Group Delay Dispersion (GDD). This oscillator design is now one of the standard ytterbium fiber laser designs used by several groups\cite{NugentGlandorf_OL2011, Li_RSI2016}. With a similar laser design, Zhou et. al. measured 28 fs pulses, the biggest difference being external prism compressors instead of grating compressors\cite{Zhou_OptExpress2008}. Both of these papers mention how difficult it is to find the right mode-locking regime that can achieve the shortest pulses, even with dispersion compensation. Work by Nugent-Glandorf et al on the RIN of these type of modelocked Yb:fiber lasers demonstrated the lowest noise is found around zero net cavity dispersion and the RIN increases as the laser moved into the normal (or anomalous) regime\cite{NugentGlandorf_OL2011}.

In published Yb:fiber laser spectra, the lasing bandwidth is smaller than the Yb emission profile, to the best of our knowledge. All of the components in the laser, including the waveplates and optical isolator, have transmission curves that aren't perfectly flat or matched, creating a complex cavity loss profile without a direct way to control or correct the spectra. Without a direct way to control the spectra, it is often an endless search of turning waveplates to switch mode-locking regimes or swapping out components and fiber. An intracavity optical filter provides a way to control the cavity loss, which will affect the optical spectra of the laser. This work investigates operation of a ytterbium-doped fiber laser with an intracavity bandpass filter to investigate the utility for achieving ultrashort pulses and the overall effect on laser performance.

Optical filters have been used in laser cavities for some time, most often to tune the wavelength, especially in fiber lased used as narrow-linewidth radiation sources. Intracavity bandpass filters have been shown to tune the wavelength for cw Yb-fiber lasers\cite{Zhang_SR2017, Mukhopadhyay_RSI2014} as well as pulsed lasers. One example is the picosecond SESAM mode-locked polarization maintaining Yb-fiber laser by Jiang et al\cite{Jiang_Optik2019}, which has 50 nm of tunability. Erbium fiber lasers share many similarities with Ytterbium and many published Erbium fiber lasers use filters of various types to achieve wavelength tunability, such as a semiconductor saturable absorber mirror (SESAM) modelocked laser with 0.9 ps pulse duration\cite{Armas_OLT2021}. Nevertheless, most pulsed fiber lasers have no explicit element for controlling the lasing wavelength.

\begin{figure}[htbp]
\includegraphics[width=8.5cm]{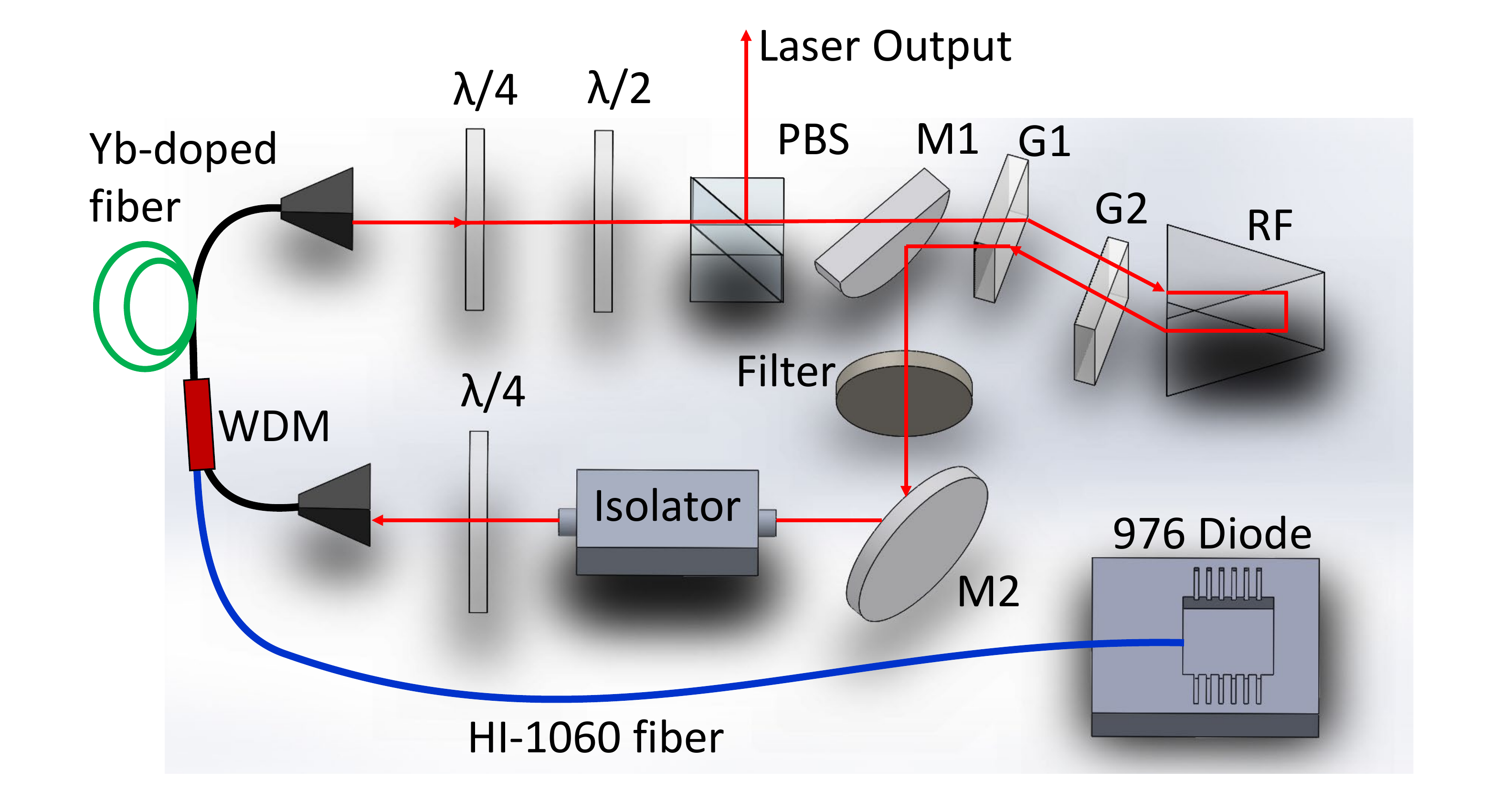}
\caption{Schematic of the Yb:fiber oscillator. Yb - Ytterbium Fiber, WDM - Wavelength Division Multiplexer, $\lambda$/2 - Half Waveplate, $\lambda$/4 - Quarter Waveplate, PBS - Polarized Beam Splitter, M - Mirror, G - Grating, RF - Retroreflecting Prism.}
\label{fig:layout2}
\end{figure}

Besides tuning the wavelength, there are other uses for filtering elements in pulsed fiber lasers, including control over the total dispersion and gain in the system. Buckley et al used a knife-edge between reflection gratings in a Yb:fiber oscillator as an optical filter with the goal of increasing the dispersion in the laser\cite{Buckley_JOSAB2007}. They were able to find stable modelocking regions when limiting the short-wavelength side of the spectrum, effectively acting as a type of long-pass filter, but reported difficulty in modelocking with the filter blocking the long-wavelength spectral region. In Mamyshev Oscillators, filters are used to obtain modelocking\cite{Olivier_OptLett2019, Rochette_IEEEPhotonTech2008}. Additionally, notch filters with only a few nanometers linewidth have also been used in all-normal-dispersion fiber lasers as a way to control the total dispersion in the lasing cavity\cite{Chong_OE2006}. Another use of optical filters in fiber laser cavities is to narrow the gain bandwidth for dissipative soliton formation\cite{Zhao_OL2010, Wang_JOSAB2017}.In 2019, Delfyett et al\cite{Delfyett_AppOpt2019} put a Waveshaper, a programmable liquid-crystal spatial light modulator, in a semiconductor frequency comb laser to explore the effect of changing the spectra on the laser characteristics. They found that the Waveshaper introduced a significant amount of phase noise and amplitude noise to the laser, although they did achieve spectral broadening. Here we investigate the application of a passive intracavity spectral filter, an optical interference filter, to increase the bandwidth of the laser, generate shorter pulses, and push the laser to operate at wavelengths outside the normal lasing bandwidth.

\section{Laser Design}

The oscillator used here is a self-similar laser\cite{Chong_RoP2015} with a sigma layout. The home-built Yb:fiber laser used in this paper is similar to those published elsewhere \cite{Li_RSI2016, NugentGlandorf_OL2011} and is shown in figure \ref{fig:layout2}. Briefly, a 500mW continuous wave 976nm diode pumps the core of a 25 cm section of Ytterbium-doped fiber. The laser cavity consists of a wavelength-division multiplexer (WDM) to combine the pump light and cavity light, 25 cm of single-mode Ytterbium-doped fiber followed by a free-space region. The non-doped sections of fiber, connecting the WDM and fiber couplers to the Yb fiber, are HI-1060 fiber. The free-space region begins with a zero-order quarter-waveplate, a zero-order half-waveplate, and a polarizing beamsplitter cube, which serves as the output coupler. The light remaining in the cavity passes through two transmission gratings (each 600 lines/mm), a retroreflecting prism, and then passes back through the gratings. The beam then goes through the optical filter, a Faraday rotator unidirectional device, and another zero-order quarter-waveplate before re-entering the fiber. The half-waveplate and two quarter waveplates along with the polarizing beamsplitter are used to control the passive modelocking through nonlinear polarization evolution (NPE) in the fiber. To achieve low-noise operation, the grating distance is set to compensate for the dispersion of the fiber such that the laser operates near net-zero dispersion.\cite{NugentGlandorf_OL2011}

\begin{figure*}
\centering \includegraphics[scale=.4]{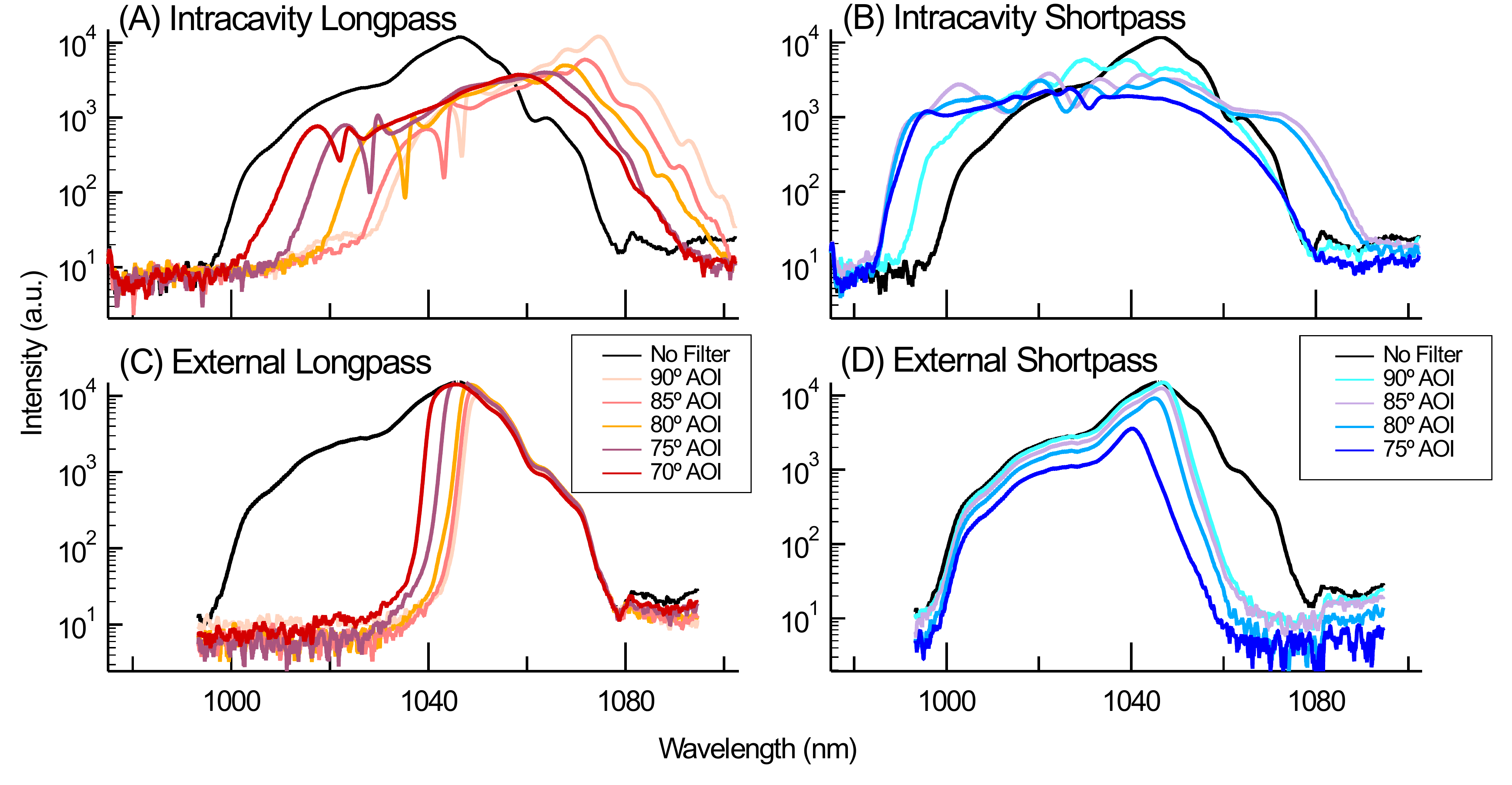}
\caption{Laser output spectra with A) longpass and B) shortpass filter placed inside the laser cavity. The filter is rotated such that the incident angle changes by 5 degree increments, as shown by the color gradients in each plot. For comparison, the C) longpass and D) shortpass filters were placed outside the cavity and angle tuned similarly.}
\label{fig:spectra}
\end{figure*} 

The laser output is separated using a 90/10 beamsplitter for compression and diagnostics, respectively. Diagnostics consist of an ASEQ high resolution B-series spectrometer and a Thorlabs PDA100A2 photodetector coupled to a Stanford Research Systems FFT, a Tetronix 1.5GHz Oscilloscope and a Rigol RF spectrum analyzer. The RIN data from 0-100 kHz was recorded with the Stanford Research Systems Fast Fourier Transform. RIN Data for higher frequencies was taken with the Rigol RF spectrum analyzer with the same photodetector. The laser output is typically 70 mW of average power at an 85 MHz repetition rate. As is documented with polarization-modelocked fiber lasers, changing waveplate positions will result in different spectra and output powers for the same grating position and pump power\cite{Leblond_PRA2002}. The data in this paper represent typical spectra that are in stable and easily-reproduced mode-locking regimes. To improve reproducibility, the waveplates were turned to minimize the RF noise of the output, which proved to be a reliable procedure to get back to the desired mode-locking regime. The pulse compression is achieved with a pair of external transmission gratings (500 lines/mm) and measured with a commercial Frequency-Resolved Optical Gating (FROG) technique, the GRENOUILLE by Swamp Optics.

A known problem of these NPE Yb:fiber oscillators is multipulsing, more than one pulse generated in the cavity. Single-pulse operation was verified using the GRENOUILLE, for very short time separations ($<$ 500 fs), the optical spectrometer for short time separations ($<$ 3 ps), and a simple homebuilt scanning autocorrelator for longer time separations. This is similar to the procedure used by one author in previous work with these oscillators \cite{Li_RSI2016}. 

Two commercially-available, stock optical interference filters were used in the laser for this work: a longpass filter (Thorlabs FELH1050) and a shortpass filter (Newport 10SWF-1050-B). The cutoff frequency of the filters can be tuned slightly by changing the incident angle of the filter. For reference, figure \ref{fig:spectra} panels C and D show the filter performance when placed outside the laser cavity and illustrate this tunability and resulting cutoff frequency. To facilitate this, the filters were placed on a graduated rotation state between the two turning mirrors in the laser cavity. The laser achieves modelocking with either the longpass or shortpass filter in the cavity with incidence angles from 90 to 70 degrees, thus allowing some tunability of the filter cutoff wavelength. Further tuning of the angle causes a significant decrease in transmission of the filter, preventing modelocking.

\section{Laser Performance}

Figure \ref{fig:spectra} compares the spectra of the filters intracavity and external to the laser cavity. At zero degrees incident angle outside the cavity, the filters have a sharp cutoff within the manufacturer tolerance of the specified 1050 nm cutoff/cut-on wavelength (3 nm for the longpass and 10 nm tolerance for shortpass). Decreasing the incident angle shifts the cutoff frequency to the blue by up to 20 nm when inside the cavity and about 5 nm when placed at the laser output. When placed inside the laser cavity, the shift is much greater due to the complex gain dynamics. Figure \ref{fig:spectra} panels A and B display the laser spectra with the intracavity filters, illustrating the spectral broadening and shifts of the centroid.  Notably, the laser shifts to new frequencies not seen in the spectra without the filter, which is what we will exploit to obtain shorter pulses.

The shortpass filter creates higher loss on the long-wavelength side of the spectra, thus pushing the spectra towards the Ytterbium absorption peak. The result is the lasing broadens both to the long and short wavelengths and results in an overall increase in the lasing bandwidth. The longpass filter doesn't significantly broaden the spectra as much as it just pushes the spectra far out to the red, beyond the original lasing bandwidth. With the longpass filter in the cavity, the small tuning of the cutoff frequency provided by rotating the filter can result in significant shifts of the centroid frequency, as shown in Figure \ref{fig:spectra} and Supplemental Information. A table of typical bandwidths is found in the Supplemental Information. Even though the overall bandwidth decreases when the longpass filter is used, there is still sufficient bandwidth for ultrafast (sub 100-fs) pulses, see Supplemental information for data.

\begin{figure}[htbp]
\centering\includegraphics[width=8cm]{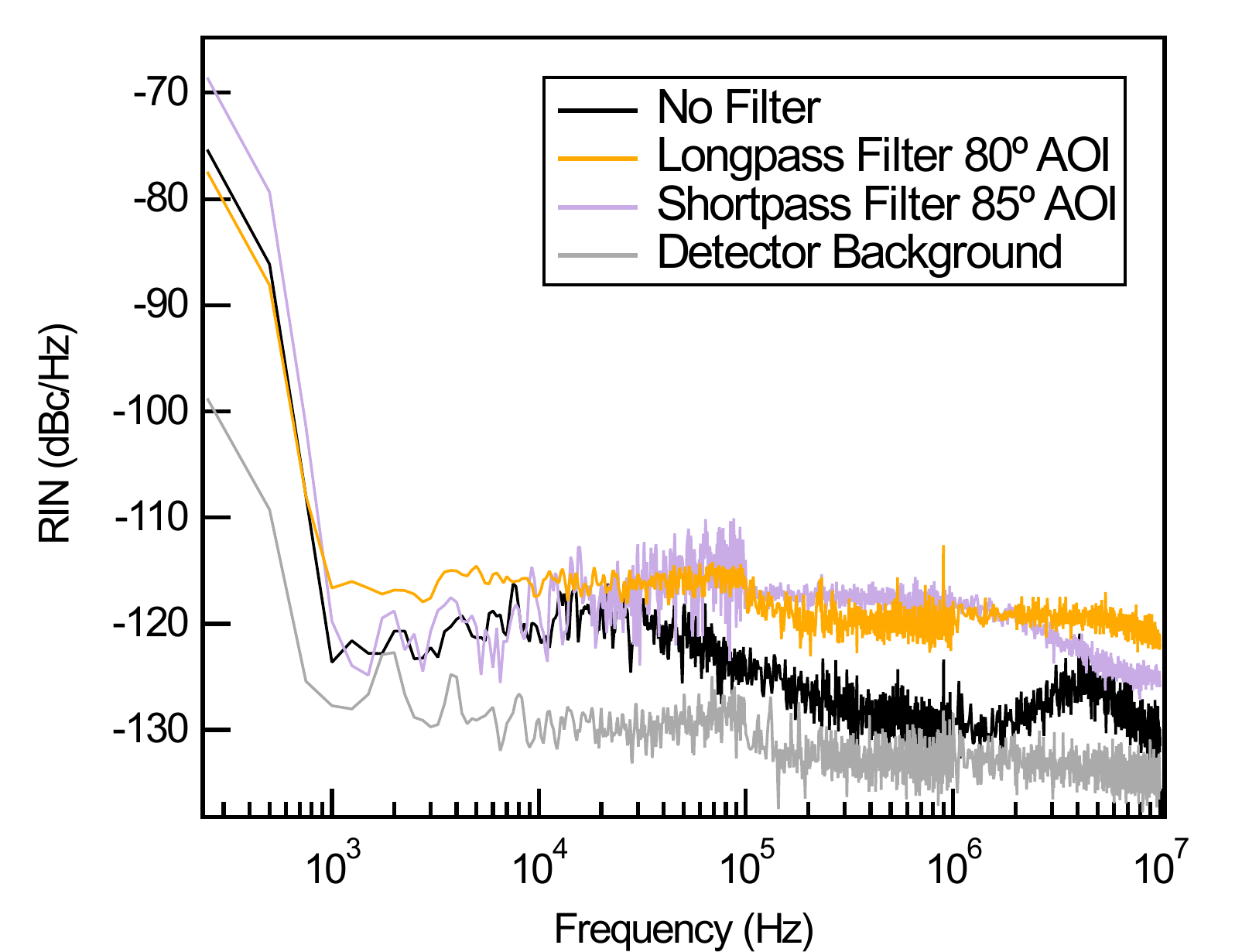}
\caption{Relative intensity noise (RIN) of the oscillator with with no filter (black), the longpass filter at 80 degress angle of incidence (yellow), the shortpass filter at 85 degrees angle of incidence (purple), and the detector background (gray).}
\label{fig:rin}
\end{figure}

To explore the effect of the filter on the laser noise, we measured the residual intensity noise (RIN) of the laser for all filter positions. Figure \ref{fig:rin} shows the RIN of the laser without a filter, with the longpass filter at 80 degrees angle of incidence, the shortpass filter at 85 degrees angle of incidence, and the noise floor of the detector. These are representative traces; the full data sets of the RIN at a range of filter positions is in the Supplemental Information. The low-frequency noise, below 20 kHz, is similar with and without the filter. At higher frequencies, the RIN with no filter is lower in noise by about 10 dB until the noise begins to drop off at 10 MHz. Overall, the laser is still a low-noise laser even with the intracavity filters.

\subsection{Pulse Compression}

\begin{figure}[htbp]
\centering\includegraphics[width=8cm]{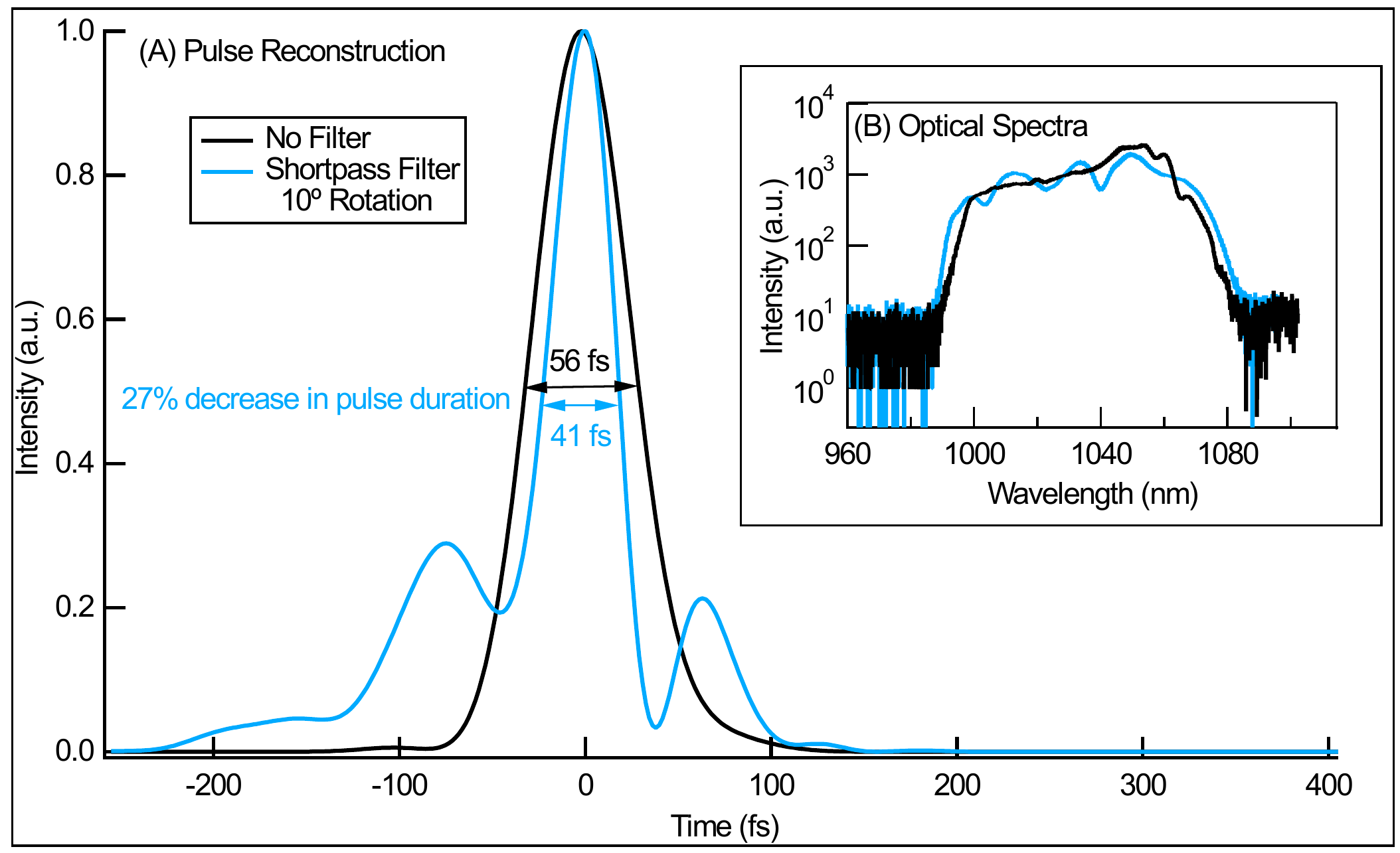}
\caption{A) Sample pulse characterisation using the shortpass filter at a 80 degree angle on incidence (blue) and the laser with no filter (black) using a commercial frequency-resolved optical gating pulse measurement (the GRENOUILLE system, Swamp Optics). B) Corresponding optical spectra }
\label{fig:frog}
\end{figure}

The filter provides a very reproducible way to achieve a broad spectra, so we explored whether this would result in a shorter pulse duration. The pulses were compressed with external compression transmission gratings and then sent to a commercial FROG (GRENOUILLE - Swamp Optics). The GRENOUILLE software reconstructs the pulse from the measured FROG trace and provides a pulse duration, among other information. Figure \ref{fig:frog} shows a pulse reconstruction and phase for a sample filter placement and for the laser without the filter. Both traces are located near the zero dispersion of the laser cavity. The minimum pulse duration without a filter in the cavity is 56 fs while the minimum pulse duration with the shortpass filter is 41.4 fs.

While we didn't achieve pulse durations as short as 36 fs\cite{ilday_OptExpress2003} or 28 fs\cite{Zhou_OptExpress2008}, we do have a method that easily positions the laser in a region to get ultrashort pulses. To achieve 36 femtosecond pulses, the authors mention that they must spend significant time turning waveplates to achieve the lowest pulse duration, which we were not able to find with our laser. The use of the filter provides a reproducible and reliable route to obtaining sub-45 fs pulses. The compression gratings will not compensate for the residual third order dispersion, so that could be limiting our measured pulse duration. 

The shortest pulses have small wings on the side. This is seen in all of our pulses as we decrease the pulse duration below about 50 fs. It was also seen in Ilday et al\cite{ilday_OptExpress2003} as they noted that 30\% of the pulse intensity is found in the wings.

\subsection{Dispersion} 

This type of Ytterbium fiber laser can operate in a variety of net cavity dispersion regimes. Nugent-Glandorf et al\cite{NugentGlandorf_OL2011} documented the effect of cavity dispersion on the RIN and comb-tooth linewidth, noting that the lowest noise was obtained near zero dispersion. Ilday et al\cite{ilday_OptExpress2003} reported their 36 fs pulse was near zero net cavity dispersion, so we systematically explored the lowest achievable pulse duration as a function of intracavity dispersion. 

The net cavity dispersion of these lasers is tuned by changing the intracavity grating distances. Using the method from Knox\cite{Knox_OL1992}, we measured the net cavity dispersion at each grating position. The cavity dispersion is changed slightly by the placement of the filter, as expected. The pulse duration as a function of dispersion for the shortpass intracavity filter is shown in Figure \ref{fig:dispersion}. Tables with the full set of data are in Supplemental Information. The laser spectra changes slightly with the dispersion and change in grating distance, so a measure of the spectral width is included as well in the Supplemental Information. The maximum spectral width is found in the slightly anomalous region, even though that didn't correspond to the shortest pulse duration. The minimum pulse duration is found near-zero net cavity dispersion, and increases on either side of zero dispersion.

\begin{figure}[htbp]
\centering\includegraphics[width=8cm]{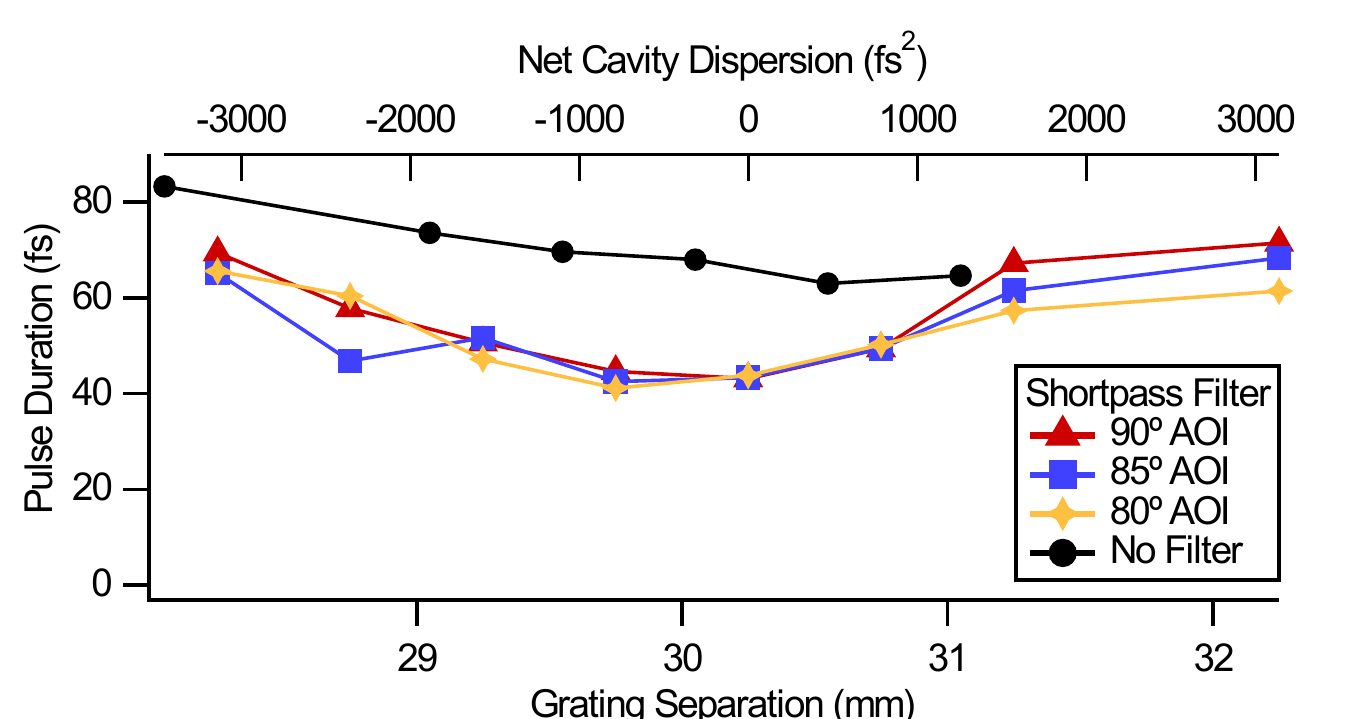}
\caption{Minimum pulse duration as a function of dispersion for the shortpass intracavity filter and without a filter.}
\label{fig:dispersion}
\end{figure}

\section{Discussion and Future Prospects}

In summary, the inclusion of a edgepass filter enables both tuning and broadening of the optical spectra of the modelocked Yb:fiber laser. Counter-intuitively, careful placement and tuning of an intracavity edgepass filter can also broaden the optical spectra beyond the original lasing spectrum. This broadened optical spectra can still be compressed into an ultrashort pulse, achieving pulses as short as 41.4 fs.  While the compressed pulses have clear wings on each side, as also reported in previous papers, and this could be detrimental for certain purposes, other uses would not be affected. For example, the use in a nonlinear process would further suppress the sidebands relative to the main pulse. While there are other methods to dropping pure Yb:fiber systems to between 20 and 65 fs\cite{Descamps_2019,Chiba_OE2015}, this method represents a simple and cheap extension to the system that requires very little expansion. While the RIN of the laser is slightly noisier with the filter in place, is is still a very low noise laser. This illustrates that the use of a simple, off-the-shelf, interference filter is useful to control the output spectrum of the laser and generate shorter pulses. A potential extension to this idea would be to use a custom-designed interference filter instead of simple shortpass and longpass filters, to achieve a more flat-topped spectra and perhaps even shorter pulses. 

We were able to significantly push the laser to the red end of the spectrum. This could be very useful if the laser is to be used in a Ytterbium amplifier. The Amplified Stimulated Emission (ASE) of these amplifiers can limit the low noise performance, by pushing the seed spectra away from the absorption peak, the ASE can be decreased.  For an ultrafast laser, one application of laser spectrum shaping is to limit gain narrowing during a subsequent amplification stage \cite{Backus_RSI1998}. Gain narrowing is the phenomenon whereby the optical spectra narrows after amplification due to the Yb gain bandwidth\cite{kuznetsova_APB2007}, which can limit the achievable pulse duration of a compressed pulse, and therefore needs to be managed to obtain ultrashort pulses. To get around this, Chiba et al found that modifying the spectra in such a way as to decrease the spectra near the gain bandwidth peak subsequently decreased the gain narrowing\cite{Chiba_OE2015}. Controlling the spectra directly out of the oscillator could be an efficient way to tune the frequency in preparation for the amplification stage.The use of the longpass filter significantly forces the optical spectra to the long wavelength region and could potentially be useful in decreasing the ASE and improving the performance of the amplifiers.

\begin{backmatter}
\bmsection{Funding} This work was funded in part by University of Georgia Startup Funding. This material is based upon work supported by the U.S. Department of Energy, Office of Science, Office of Basic Energy Sciences, Gas Phase Chemical Physics Program under Award Number DE-SC0020268. This work is also funded by the American Chemical Society Petroleum Research Fund grant 59750-DNI6. This work was funded in part by an award from the University of Georgia Innovation Gateway through its IP Development Program. U.M.T would like to that University of Georgia Department of Chemistry, Summer Research Opportunity (SURO) grant for funding.

%\bmsection{Disclosures} The authors declare no conflicts of interest.

%\bmsection{Data Availability} Data presented in this paper are available from the corresponding author upon reasonable request.

\end{backmatter}

% Bibliography
\bibliography{References}

% Full bibliography added automatically for Optics Letters submissions; the following line will simply be ignored if submitting to other journals.
% Note that this extra page will not count against page length
\bibliographyfullrefs{References}

\end{document}